\documentclass[twocolumn]{revtex4-1}
\usepackage{graphicx,amsmath,amssymb,amsfonts}   
\usepackage{dcolumn}    
\usepackage{bm}         
\usepackage{color}

\newcommand{\mb}[1]{\mathbf{#1}}
\newcommand{\ie}{\emph{i.e.\@} }
\newcommand{\etal}{\emph{et al.\@}}\newcommand{\diffd}{\text{d}}

\begin{document}
\title{Degenerate transition pathways for screw dislocations: implications for migration}
\author{M. R. Gilbert}
\author{S. L. Dudarev}
\affiliation{CCFE/EURATOM Fusion Association, Culham Science Centre, \\Abingdon, Oxfordshire, OX14 3DB, UK}
\author{P. M. Derlet}
\affiliation{Condensed Matter Theory Group, Paul Scherrer Institute, CH-5232 Villigen PSI, Switzerland}

\date{November 2001}

\begin{abstract}
In body-centred-cubic (bcc) metals migrating  \(1/2\langle 111\rangle\) screw dislocations experience a periodic energy landscape with a triangular symmetry. Atomistic simulations, such as those performed using the nudged-elastic-band (NEB) method, generally predict a transition-pathway energy-barrier with a double-hump; contradicting {\it{Ab Initio}} findings. Examining the trajectories predicted by NEB for a particle in a Peierls energy landscape representative of that obtained for a screw dislocation, reveals an unphysical anomaly caused by the occurrence of monkey saddles in the landscape. The implications for motion of screws with and without stress are discussed.
\\\\\noindent Note: the present version of this manuscript was written in 2011 and does not reflect the current understanding of the Peierls landscape. In particular, it is now thought that, for Fe and other bcc transition metals, the saddle-point or ``split-core'' configuration is in fact an energy maximum~\cite{itakuraetal2012,ventelonetal2013}.
\end{abstract}

\maketitle

In body-centred-cubic (bcc) transition metals, such as iron and tungsten, \(1/2[111]\) screw dislocations play a critical role in plastic deformation. This is particularly important in irradiated materials, where the irradiation damage may alter the mobility of screw dislocations, leading to radiation induced hardening and embrittlement.

Consequently, it is vitally important to understand the mechanisms and processes involved in the motion of screw dislocations. It is almost impossible to observe, let alone investigate, screw dislocations in experiment, and so computational simulation at the atomic level has a crucial role to play.

One of the main avenues for investigating screw dislocations via simulation is through the use of interatomic potentials in molecular dynamics. Density functional theory (DFT) calculations have revealed both the 0K core structure and 0K migration barrier for \(1/2[111]\) screw dislocations (see for example~\cite{ventelonetal2007}) and the fitting of potentials is now directed toward reproducing these properties. However, while it is generally understood how to produce the correct non-generate compact core structure~\cite{gilbertdudarev2010}, it is less clear how to obtain the Peierls migration barrier for a dislocation moving between adjacent equilibrium (`easy') core positions. DFT calculations predict a single-hump barrier, while atomistic calculations for potentials predicting the correct core structure, either via the nudged-elastic-band (NEB)~\cite{gilbertdudarev2010} or drag method~\cite{ventelonetal2007}, find a double-hump structure with a metastable intermediate configuration.

In this letter, we investigate the origin of the double-hump structure predicted by atomistic simulations and suggest an explanation for why it occurs. This in turn leads to the revelation that interatomic potentials may, in fact, be modelling the migration barrier correctly. Furthermore, the consequences for the migration of screw dislocations are profound, with reasons behind alternative glide planes -- for so-long a contentious issue -- becoming immediately understandable from analysis of the energy landscape.

\begin{figure}[h]
\frame
{\includegraphics[trim=1cm 4.5cm 0.5cm 4cm,clip=true,angle=-90,width=0.4\textwidth]{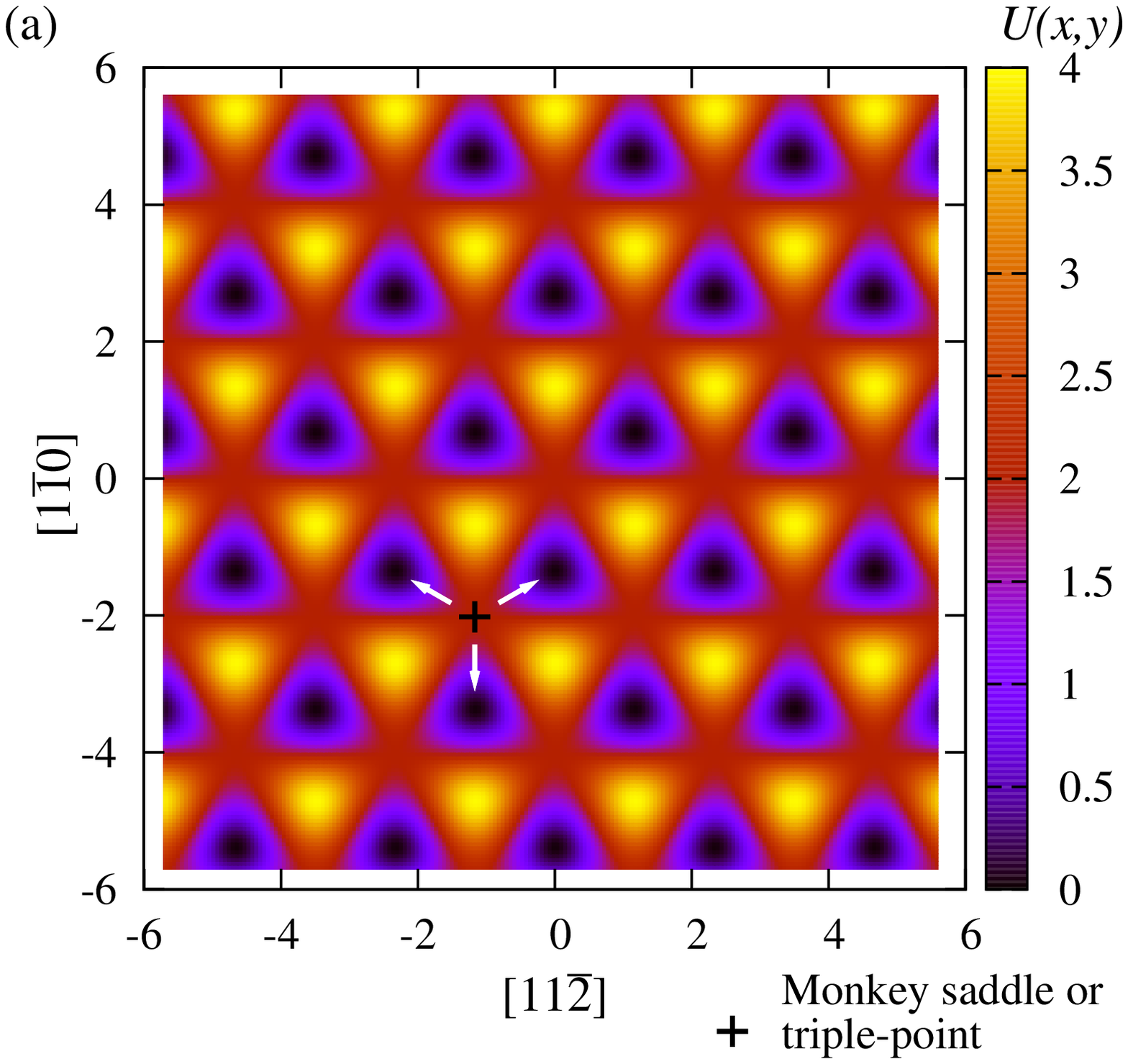}}
\frame
{\includegraphics[trim=1cm 0cm 0cm 0cm,clip=true,angle=-90,width=0.4\textwidth]{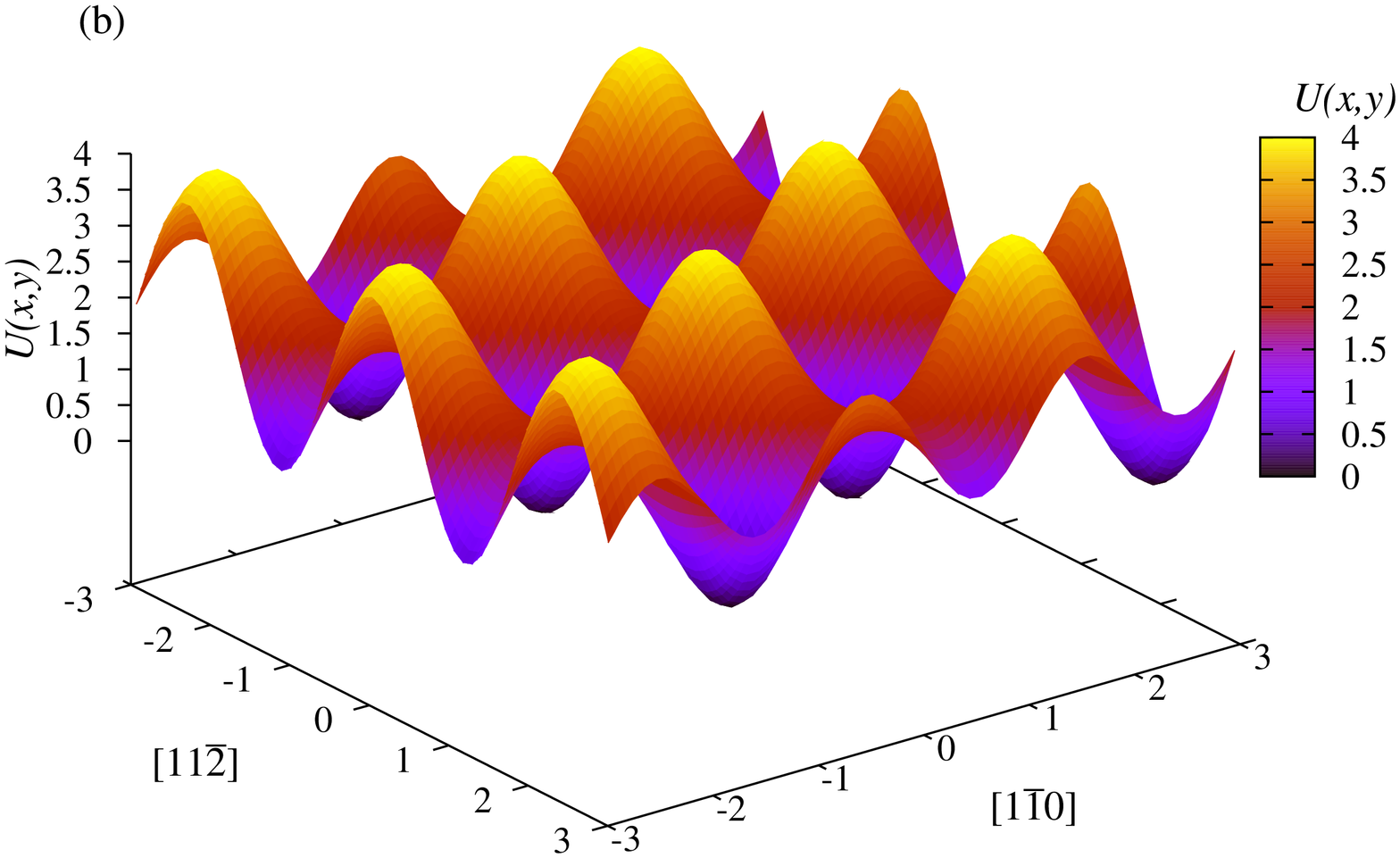}}
\caption{\label{edagawa_pot}The energy landscape defined by the Edagawa~\cite{edagawaetal1997} potential for a point-like screw dislocation (equation~\eqref{edagawa1}) with \(V_{\text{mag}}=2\) and \(a=2.8553\)~{\AA} --  the lattice parameter of the Mendelev~\cite{mendelevhan2003} potential for Fe; (a) as a two-dimensional contour plot for the variation in energy, and (b) as three-dimensional energy landscape with the contours super-imposed onto the energy surface.}
\end{figure}

It is well known that the displacements associated with a \(1/2\langle 111\rangle]\) screw dislocation are essentially one-dimensional in nature and parallel to the Burgers vector (Clouet \etal~\cite{clouetventelon2009} show that the displacements perpendicular to the Burgers vector are at least an order of magnitude smaller than those parallel to it). Previously, this has allowed the core structure of screws to be investigate by considering the bcc lattice as being constructed of rigid [111] atomic strings in a Multi-String Frenkel-Kontorova (MSFK) model~\cite{gilbertdudarev2010}. However, if we do not care about the core structure (other than to assume it is correct), then it is not necessary to consider the interaction of strings. Instead a screw dislocation can be reduced to a single point defined by its position in the 2D space perpendicular to its Burgers vector.
Edagawa \etal~\cite{edagawaetal1997} defined a sinusoidal potential for the energy of such a \emph{screw-particle}:
\begin{align}\label{edagawa1}
V(x,y)=&\frac{8}{3\sqrt{3}}V_{\text{mag}}\sin\left[\sqrt{2}\pi\frac{y}{a}\right]\cdot\\*\nonumber
&\sin\left[\frac{\pi}{\sqrt{2}a}(y+\sqrt{3}x)\right]\sin\left[\frac{\pi}{\sqrt{2}a}(y-\sqrt{3}x)\right],
\end{align}
where \(a\) is the usual lattice parameter. In this formalism, x is \([11\bar{2}]\) y is \([1\bar{1}0]\), and the screw-particle has \(-V_{\text{mag}}\) energy minima representing the easy-cores
and \(V_{\text{mag}}\) hard-core maxima correctly distributed in the triangular lattice of a (111) plane (Fig.~\ref{edagawa_pot}).

\begin{figure}[b]
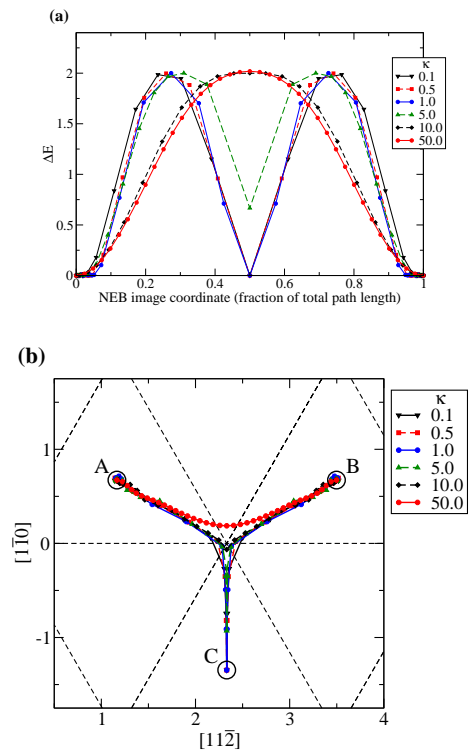

{\includegraphics[width=0.3\textwidth,trim=0cm 0cm 0cm 0cm,clip=true]
{neb_point_barriers_39_0_red.eps}}
\\\vspace{0.5cm}
{\includegraphics[height=0.3\textwidth,trim=0cm 0cm 0cm 0cm,clip=true]
{neb_point_trajectories_39_0_red.eps} }
\caption{\label{point_NEBs}NEB trajectories as a function of the spring constant \(\kappa\) for a screw-particle moving in the \([11\bar{2}]\) direction between two adjacent energy minima (from A to B in (b)) of the Edagawa potential~\eqref{edagawa1}. (a) the energy barriers associated with trajectories, and (b) the paths taken by the particle in the (111) plane as it moves from A to B. For sufficiently small values of \(\kappa\) the particle moves on a path that also includes the minima C, which is the third `easy' position surrounding the saddle point.}
\end{figure}

Using this potential static NEB calculations\cite{millsjonsson1994,jonssonetal1998} give the Peierls-energy barrier associated with the translation of a particle from one energy minimum to another adjacent to it (from A to B in Fig.~\ref{point_NEBs}b). Fig.~\ref{point_NEBs}a shows a series of such calculations for different values of the NEB spring constant \(\kappa\) for a screw-particle moving the \(2a/\sqrt{6}\)~{\AA} in a \([112]\) direction, while Fig.~\ref{point_NEBs}b shows the pathway of the particle in 2D space for each of the trajectories.
Note that the variation as a function of \(\kappa\) is reproduced regardless of the particular tangent method used in the NEB method, even if using ones designed to ensure a smooth trajectory (see for example Henkelman \etal~\cite{henkelmanjonsson2000}).

Trajectories with single-humped energy barriers are characterised by a relatively smooth arcing pathway between the two endpoints, while double-humped barriers result from trajectories that deviate significantly from this, with the mid-point particle position approaching the third easy-core minima adjacent to the `saddle point' between the trajectory endpoints (C in Fig.~\ref{point_NEBs}b). This three-way saddle-point  in the Peierls energy landscape of the Edagawa potential~\eqref{edagawa1} is also known as a\emph{monkey saddle}.

\begin{figure}[t]
{\includegraphics[height=0.4\textwidth,trim=1cm 0.4cm 0cm 1cm,clip=true,angle=-90]{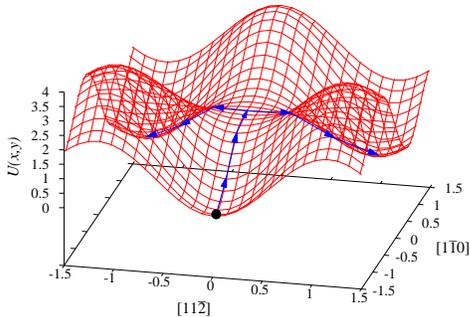}}
\caption{\label{3D_mountain_profile}The energy landscape of the Edagawa potential~\eqref{edagawa1} showing the two alternative forward directions possible for a screw-particle that has moved from an energy minimum to the top of the monkey saddle.}
\end{figure}

A particle (or indeed a screw dislocation in a similar energy landscape) sitting at the top of this monkey saddle can follow three equally-likely downward paths, as demonstrated by the (white) arrows for the highlighted monkey saddle in Fig.~\ref{edagawa_pot}a, and, correspondingly, a particle travelling up to this saddle from one of three energy minima surrounding it has two equally-favourable forward alternatives when it reaches the top of the barrier. Thus the screw-particle has a frustrated forward trajectory (Fig.~\ref{3D_mountain_profile}).

Clearly, a single hump is the most physical result since it is a true representation of the barrier between adjacent minima. The double hump, on the other hand, appears to be an artefact of the NEB method because it only appears when \(\kappa\) is sufficiently small, \ie when the springs between image configurations along the reaction coordinate are sufficiently weak. For a true saddle point with only two options at the top of the barrier the NEB method has no problem, but the unique nature of the monkey saddle causes un-physical trajectories to be accepted simply because they correspond to a lower overall barrier energy (summed over the energy of the discrete image configurations).

For the present situation, starting from the single-humped barrier corresponding to a simple linear path between two low-energy positions, the NEB algorithm will search for a lower energy trajectory by perturbing the images from their current positions in the direction of downward force gradients. For most images these gradients are along the trajectory (\ie toward the minima at each endpoint), and are projected out so as to prevent the images slipping towards the endpoints, which would otherwise cause the barrier to be under-represented. However, at the monkey saddle there is also a downwards gradient normal to the trajectory towards the third energy minima surrounding the saddle. These forces are not projected out, and therefore produce a tendency to perturb an image at the top of the saddle towards the third minima. If the springs are sufficiently strong then the image configuration is held effectively at the monkey saddle, but if \(\kappa\) is small then the trajectory slips away from the saddle and a double-hump energy barrier results.  If it is not known \emph{a priori} what form a particular energy barrier should take and if the overall energy landscape is not well understood (as in the case of a \(1/2\langle 111\rangle\) screw dislocation), then it is conceivable that such an artificial barrier might be believed true.

\begin{figure}[t]
{\includegraphics[height=0.4\textwidth,trim=1cm 4.5cm 0.5cm 4cm,clip=true,angle=-90]{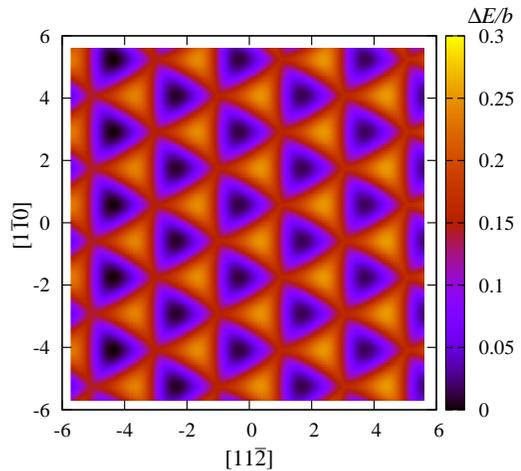}}
\caption{\label{mend_pot}2D representation of the Peierls-potential landscape for a \(1/2\langle 111\rangle\) screw dislocation under the Mendelev~\cite{mendelevhan2003} for Fe.}
\end{figure}

Knowing the Peierls energy landscape for a particular transition, as we do here for a screw-particle in the Edagawa potential (Fig.~\ref{edagawa_pot}), and observing the monkey saddles it contains, makes it immediately clear that a single-humped transition energy barrier is the correct one. To confirm the existence of the monkey saddle in the energy landscape of a \(1/2\langle 111\rangle\) screw dislocation generated from atomistic simulations we start by assuming that the isotropic elasticity solution for the screw is a good representation of its relaxed structure, at least for the Mendelev~\cite{mendelevhan2003} potential in Fe considered here. Gilbert and Dudarev~\cite{gilbertdudarev2010} observed that this was true for the fully relaxed easy-core configuration. Thus we compute the energy landscape by measuring the instantaneous energy of a screw-dislocation 20 Burgers vectors long, described by the isotropic solution, as a function of its position in the (111) plane, and obtain the result shown in Fig.~\ref{mend_pot}.

The results demonstrate that, at least for the Mendelev potential of Fe, the Peierls energy landscape experienced by a perfect screw dislocation does contain the same monkey saddles as those seen in the landscape defined by the Edagawa potential~\eqref{edagawa1}. In particular, there is nothing in Fig.~\ref{mend_pot} to suggest that the double-humped translation barrier previously predicted for the Mendelev potential~\cite{ventelonetal2007} is representative of the trajectory a screw would take through this landscape -- there are no metastable configurations that could account for the dip in the middle of the barrier.

\begin{figure}[t]
{\includegraphics[width=0.4\textwidth,trim=0cm 0cm 0cm 0cm,clip=true]{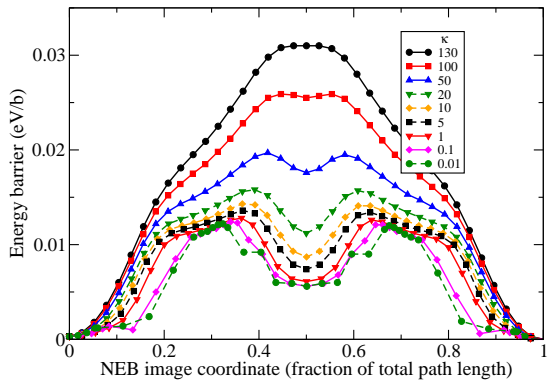}}
\caption{\label{mendelev_neb}A series of NEB trajectories for the translation of a screw dislocation from one Peierls valley to the next in a \([11\bar{2}]\) direction under the Mendelev~\cite{mendelevhan2003} interatomic potential for Fe. The only difference between each curve is the size (strength) of the spring constant \(\kappa\) between each image along the trajectory.}
\end{figure}

Given this new understanding we can investigate whether the true migration barrier can be reproduced for atomistic systems using the NEB technique by simply varying~\(\kappa\). NEB calculations have been performed on a \(1/2\langle 111\rangle\) screw-dislocation dipole translating in the \((\bar{1}10)\) plane (in opposite \([112]\) directions). The two dislocation poles were placed 20 \([\bar{1}10]\) unit-cell dimensions apart (\ie \(20\times\sqrt{2}a\)) in a bcc lattice containing 12000 atoms of size \(10\times 40\times 5\) unit cells in a \(\{[11\bar{2}],[\bar{1}10],[111]\}\) coordinate system. Periodic boundary conditions (PBCs) were used in all directions. For each NEB calculation 37 images were used along the trajectory, giving 39 configurations in total for each curve in Fig.~\ref{mendelev_neb}.

The results (Fig.~\ref{mendelev_neb}) indicate that while increasing the value of \(\kappa\) does eventually produce a single-humped energy barrier, the height of the barrier does not converge, but instead continues to increase, and so  it may be that the NEB method (or indeed the drag method) is fundamentally unsuitable to measure Peierls barriers in energy landscapes containing monkey saddles.

The frustration that results from the presence of monkey saddles in the energy landscape of a screw dislocation can be appreciated by considering the diffusion at finite temperature of a screw-particle through the Edagawa potential~\eqref{edagawa1}. Starting from the over-damped (meaning that acceleration terms are neglected) equation of motion for particle in a 2D potential \(U(\mb{x})\):
\begin{equation}\label{eq_of_mot}
\gamma\frac{\diffd\mb{x}}{\diffd t}=-\frac{\diffd U}{\diffd\mb{x}}+\mb{F}(t),
\end{equation}
where \(\mb{x}=(x,y)\), \(\gamma\) is the friction coefficient, and \(\mb{F}(t)\) is a delta-correlated random force vector with each component satisfying
\begin{equation}\label{eq_of_mot_cond}
\langle F_k(t)F_k(t')\rangle=\Gamma^2\delta(t-t'), \text{ and } \langle F_k(t)\rangle=0
\end{equation}
Combining the theories of Einstein~\cite{einstein1905} and Langevin~\cite{langevin1908} for Brownian motion, we have that~\cite{coffeyetal2004}
\begin{align}
\gamma=&\frac{k_BT}{D}, \text{  and so}\\
\Gamma=&\sqrt{2D}k_bT,
\end{align}
by substituting, into equation \eqref{eq_of_mot_cond}, the solution to equation~\eqref{eq_of_mot} in the absence of an external stress (\ie  \(U\equiv 0\))~\cite{derletmanhdudarev2007,dudarevgilbertetal2010}. Here \(D\) is the diffusion coefficient of the particle, \(k_B\) is Boltzmann's constant, and \(T\) is temperature.
Following the arguments by Frenkel and Smit~\cite{frenkelsmit2002} for non-commutative Liouville operators (\ie \(\diffd U/\diffd\mb{x}\) and \(\mb{F}(t)\)) we evolve each coordinate \(k\) of the particle's position by \(\Delta t\) in three steps:
\begin{align}
&\text{i)   }&x_k^*&=x_k(t)+\frac{\Delta t}{2}F_k(t);\\
&\text{ii)  }&x_k^{**}&=x_k^*-\frac{\Delta t}{\gamma}\frac{\partial U(x^*_k)}{\partial x_k};\\
&\text{iii) }&x_k(t+\Delta t)&=x_k^{**}+\frac{\Delta t}{2}F_k(t).
\end{align}
In steps (i) and (iii), \(1/\gamma\) has been absorbed into \(F_k(t)=\xi_k\sqrt{2D\Delta t}\), where \(\xi_k\) is a Gaussian-distributed random number between -1 and 1, introduced in order to satisfy \eqref{eq_of_mot_cond}, and which is fixed in each time step for each coordinate direction.

\begin{figure}[t]
{\includegraphics[width=0.4\textwidth,trim=0cm 4cm 0.5cm 3cm,clip=true,angle=-90]{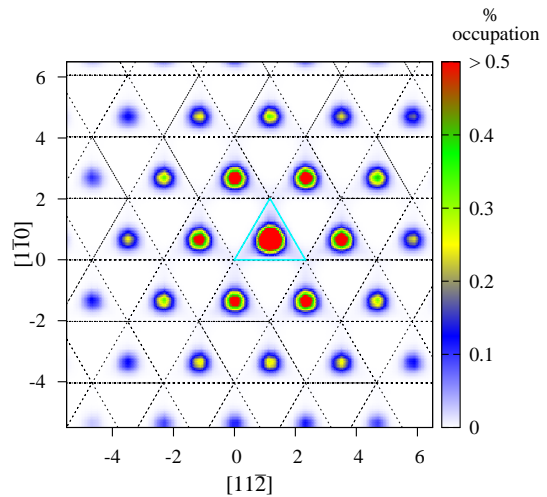}}
\caption{\label{point_diffusion_trajectories} a contour plot of the \% occupation of the landscape over the course of 1000 20~ps trajectories of a screw-particle migrating through the Edagawa energy-landscape. Each trajectory starts from the centre of the highlighted easy-core triangle, and the only difference between trajectories is the initial random number seed.}
\end{figure}

In Fig.~\ref{point_diffusion_trajectories}, the percentage occupation over the course of one thousand 20~ps trajectories has been calculated for a screw-particle moving in the Edagawa potential at a temperature \(T\) of 2000K, and under a friction coefficient chosen to produce a diffusion coefficient \(D\) of \(4\times 10^{12}\)~{\AA}\(^2\)s\(^{-1}\)(the temperature and diffusion coefficient are very high compared to experiment and atomistic simulations, but were used for illustration only), with \(a=2.8553\), and \(V_{\text{mag}}=2\). In each trajectory the screw-particle was initially at the centre of the easy-core triangle high-lighted in bold (light-blue) in the figure.

Predictably, the region with the highest occupation is at the centre of the easy-core triangle from which each trajectory was initiated, and the total occupation of points within this region are of the order of 3\%. The reduction in occupation in the first six triangles around the initial position is uniform in all directions, but beyond this the distribution of occupations becomes non-uniform as a result of the frustration experienced at the monkey saddles.  \(\langle 110\rangle\) directions are more favourable than \(\langle 112\rangle\) because there are twice as many equally probable shortest routes to each triangle in the former direction compared to the latter. Of course, as the length of trajectories increases, leading to spreading over a greater range of potential minima, this asymmetry would become less and less obvious because of the greater number of routes of equal length to any given energy well.

\begin{figure}[t]
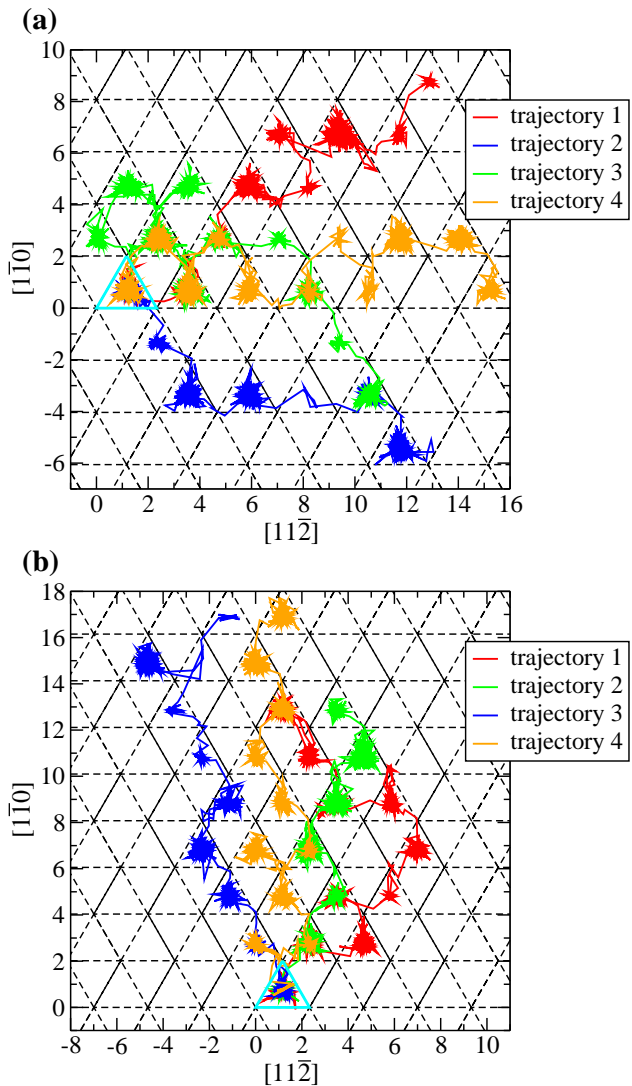

{\includegraphics[height=0.4\textwidth,trim=0cm 0cm 0cm 0cm,clip=true]{neb_point_diffusion_trajectories_biased112_20ps.eps}}
{\includegraphics[height=0.4\textwidth,trim=0cm 0cm 0cm 0cm,clip=true]{neb_point_diffusion_trajectories_biased110_20ps.eps}}
\caption{\label{point_diffusion_trajectories_biased}The effect of a bias to the Edagawa potential landscape on the trajectory of the screw-particle; (a) four trajectories experiencing a negative energy slope in the \([11\bar{2}]\) direction, and (b) four trajectories biased by a negative gradient in the \([1\bar{1}0]\) direction. Starting core position highlighted in bold (light-blue).}
\end{figure}

On the other hand, if there is any bias in the energy landscape, perhaps due to an external stress field then the picture can be altered dramatically. Figs.~\ref{point_diffusion_trajectories_biased}a and~\ref{point_diffusion_trajectories_biased}b both show four trajectories that result from adding a negative potential gradient to equation~\eqref{edagawa1} in the \([11\bar{2}]\) and \([1\bar{1}0]\) directions, respectively. In each case this results in the trajectories being strongly biased to move in the direction of the potential gradient.

In a real system, \ie one containing a long screw dislocation segment, it is well known that the dislocation moves by forming kinks, which can, once formed, drag the rest of the dislocation over the potential barrier. The above observations would suggest that this kink formation process is frustrated by the monkey saddle configuration. The fact that returning to the original position is only half as likely as being displaced provides a possible explanation for the observed discrepancy between experiment and simulation. Experiments generally predict an activation energy for screw dislocations that is roughly half of that predicted by molecular dynamics simulations. While simulations of screw dislocations are generally performed in a landscape heavily biased by large stresses and strains, which would tend to remove the frustration (see Fig.~\ref{point_diffusion_trajectories_biased}), experiments are normally performed at much lower stresses (or with no stress at all). In such experiments, the frustration would be fully realised, leading to propagation that is twice as likely, and therefore has only half the activation energy, as that measured by simulation.

In summary, the simulations of a screw-particle in the sinusoidal Edagawa~\cite{edagawaetal1997} potential (equation~\eqref{edagawa1}) reveal that the nudged-elastic-band (NEB) method produces anomalous double-humped Peierls-energy barriers for the migration of the particle through the landscape. This is caused by the presence of a monkey saddle-point in the energy landscape, which has equal positive forces pointing towards each of three easy-core energy minima, leading conditions whereby the pathway can anomalously deviate toward a third easy-core (different from the endpoints of the pathway).

Atomistic simulations for existing interatomic potentials have shown that the energy landscape for a migrating screw dislocation also contains these three-way saddle points, and so the NEB method will also converge to the dynamically unphysical double-hump Peierls-energy barrier. In a real system, if a screw dislocation (or screw-particle) had actually started to move towards this third minima it would, in all likelihood, continue to the bottom - a complete transition over the saddle-point with a single-humped energy barrier. The double-humped barrier previously predicted for the motion of \(1/2\langle 111\rangle\) screw dislocations in bcc metals is actually two transitions over the saddle-point, while the height and shape of each single-hump is the true picture for a single transition from one easy-core minimum to the next in one of the three \(\langle 112\rangle\) directions in the \((111)\) plane.

A further consequence of the observed monkey saddle configuration is that propagation of a screw dislocation through a lattice (either via kinks or as a straight-line) is twice as likely as remaining in the original core position. If a dislocation segment propagates to the top of a monkey saddle (perhaps as a result of thermal fluctuations), then there are two forward routes and a return to the original position that are all equally favourable. Thus at finite temperature net movement away from the initial position would happen twice as often as in a system containing only two-way saddles in the Peierls-energy landscape.

Hence, a possible explanation for why experimental measurements of the activation energy for the screw-migration is half of that predicted from simulation is that experiment generally measure the activation energy  in an unbiased system of monkey saddles where the frequency of motion (displacement) is twice that of a dynamic process containing normal saddle points. Meanwhile, simulations are able to explicitly calculate the the energy associated with an individual transition of a screw dislocation over the barrier.

We gratefully acknowledge stimulating and helpful discussion with J. Marian. This work was funded by the RCUK Energy Programme under grant EP/I501045 and the European Communities under the contract of Association between EURATOM and CCFE. The views and opinions expressed herein do not necessarily reflect those of the European Commission. This work was carried out within the framework of the European Fusion Development Agreement.
\bibliography{peierls_paper}

\begin{thebibliography}{16}%
\makeatletter
\providecommand \@ifxundefined [1]{%
 \@ifx{#1\undefined}
}%
\providecommand \@ifnum [1]{%
 \ifnum #1\expandafter \@firstoftwo
 \else \expandafter \@secondoftwo
 \fi
}%
\providecommand \@ifx [1]{%
 \ifx #1\expandafter \@firstoftwo
 \else \expandafter \@secondoftwo
 \fi
}%
\providecommand \natexlab [1]{#1}%
\providecommand \enquote  [1]{``#1''}%
\providecommand \bibnamefont  [1]{#1}%
\providecommand \bibfnamefont [1]{#1}%
\providecommand \citenamefont [1]{#1}%
\providecommand \href@noop [0]{\@secondoftwo}%
\providecommand \href [0]{\begingroup \@sanitize@url \@href}%
\providecommand \@href[1]{\@@startlink{#1}\@@href}%
\providecommand \@@href[1]{\endgroup#1\@@endlink}%
\providecommand \@sanitize@url [0]{\catcode `\\12\catcode `\$12\catcode
  `\&12\catcode `\#12\catcode `\^12\catcode `\_12\catcode `\%12\relax}%
\providecommand \@@startlink[1]{}%
\providecommand \@@endlink[0]{}%
\providecommand \url  [0]{\begingroup\@sanitize@url \@url }%
\providecommand \@url [1]{\endgroup\@href {#1}{\urlprefix }}%
\providecommand \urlprefix  [0]{URL }%
\providecommand \Eprint [0]{\href }%
\providecommand \doibase [0]{http://dx.doi.org/}%
\providecommand \selectlanguage [0]{\@gobble}%
\providecommand \bibinfo  [0]{\@secondoftwo}%
\providecommand \bibfield  [0]{\@secondoftwo}%
\providecommand \translation [1]{[#1]}%
\providecommand \BibitemOpen [0]{}%
\providecommand \bibitemStop [0]{}%
\providecommand \bibitemNoStop [0]{.\EOS\space}%
\providecommand \EOS [0]{\spacefactor3000\relax}%
\providecommand \BibitemShut  [1]{\csname bibitem#1\endcsname}%
\let\auto@bib@innerbib\@empty
\bibitem [{\citenamefont {Itakura}\ \emph {et~al.}(2012)\citenamefont
  {Itakura}, \citenamefont {Karburaki},\ and\ \citenamefont
  {Yamaguchi}}]{itakuraetal2012}%
  \BibitemOpen
  \bibfield  {author} {\bibinfo {author} {\bibfnamefont {M.}~\bibnamefont
  {Itakura}}, \bibinfo {author} {\bibfnamefont {H.}~\bibnamefont {Karburaki}},
  \ and\ \bibinfo {author} {\bibfnamefont {M.}~\bibnamefont {Yamaguchi}},\
  }\href@noop {} {\bibfield  {journal} {\bibinfo  {journal} {Acta Mater.}\
  }\textbf {\bibinfo {volume} {60}},\ \bibinfo {pages} {3698} (\bibinfo {year}
  {2012})}\BibitemShut {NoStop}%
\bibitem [{\citenamefont {Ventelon}\ \emph {et~al.}(2013)\citenamefont
  {Ventelon}, \citenamefont {Willaime}, \citenamefont {Clouet},\ and\
  \citenamefont {Rodney}}]{ventelonetal2013}%
  \BibitemOpen
  \bibfield  {author} {\bibinfo {author} {\bibfnamefont {L.}~\bibnamefont
  {Ventelon}}, \bibinfo {author} {\bibfnamefont {F.}~\bibnamefont {Willaime}},
  \bibinfo {author} {\bibfnamefont {E.}~\bibnamefont {Clouet}}, \ and\ \bibinfo
  {author} {\bibfnamefont {D.}~\bibnamefont {Rodney}},\ }\href@noop {}
  {\bibfield  {journal} {\bibinfo  {journal} {Acta Mater.}\ }\textbf {\bibinfo
  {volume} {61}},\ \bibinfo {pages} {3973} (\bibinfo {year}
  {2013})}\BibitemShut {NoStop}%
\bibitem [{\citenamefont {Ventelon}\ and\ \citenamefont
  {Willaime}(2007)}]{ventelonetal2007}%
  \BibitemOpen
  \bibfield  {author} {\bibinfo {author} {\bibfnamefont {L.}~\bibnamefont
  {Ventelon}}\ and\ \bibinfo {author} {\bibfnamefont {F.}~\bibnamefont
  {Willaime}},\ }\href@noop {} {\bibfield  {journal} {\bibinfo  {journal} {J.
  Computer-Aided Mater. Des.}\ }\textbf {\bibinfo {volume} {14}},\ \bibinfo
  {pages} {85} (\bibinfo {year} {2007})}\BibitemShut {NoStop}%
\bibitem [{\citenamefont {Gilbert}\ and\ \citenamefont
  {Dudarev}(2010)}]{gilbertdudarev2010}%
  \BibitemOpen
  \bibfield  {author} {\bibinfo {author} {\bibfnamefont {M.~R.}\ \bibnamefont
  {Gilbert}}\ and\ \bibinfo {author} {\bibfnamefont {S.~L.}\ \bibnamefont
  {Dudarev}},\ }\href@noop {} {\bibfield  {journal} {\bibinfo  {journal}
  {Philos. Mag.}\ }\textbf {\bibinfo {volume} {90}},\ \bibinfo {pages} {1035}
  (\bibinfo {year} {2010})}\BibitemShut {NoStop}%
\bibitem [{\citenamefont {Clouet}\ \emph {et~al.}(2009)\citenamefont {Clouet},
  \citenamefont {Ventelon},\ and\ \citenamefont
  {Willaime}}]{clouetventelon2009}%
  \BibitemOpen
  \bibfield  {author} {\bibinfo {author} {\bibfnamefont {E.}~\bibnamefont
  {Clouet}}, \bibinfo {author} {\bibfnamefont {L.}~\bibnamefont {Ventelon}}, \
  and\ \bibinfo {author} {\bibfnamefont {F.}~\bibnamefont {Willaime}},\
  }\href@noop {} {\bibfield  {journal} {\bibinfo  {journal} {Phys. Rev. Lett.}\
  }\textbf {\bibinfo {volume} {102}},\ \bibinfo {pages} {055502} (\bibinfo
  {year} {2009})}\BibitemShut {NoStop}%
\bibitem [{\citenamefont {Edagawa}\ \emph {et~al.}(1997)\citenamefont
  {Edagawa}, \citenamefont {Suzuki},\ and\ \citenamefont
  {Takeuchi}}]{edagawaetal1997}%
  \BibitemOpen
  \bibfield  {author} {\bibinfo {author} {\bibfnamefont {K.}~\bibnamefont
  {Edagawa}}, \bibinfo {author} {\bibfnamefont {T.}~\bibnamefont {Suzuki}}, \
  and\ \bibinfo {author} {\bibfnamefont {S.}~\bibnamefont {Takeuchi}},\
  }\href@noop {} {\bibfield  {journal} {\bibinfo  {journal} {Phys. Rev. B}\
  }\textbf {\bibinfo {volume} {55}},\ \bibinfo {pages} {6180} (\bibinfo {year}
  {1997})}\BibitemShut {NoStop}%
\bibitem [{\citenamefont {Mendelev}\ \emph {et~al.}(2003)\citenamefont
  {Mendelev}, \citenamefont {Han}, \citenamefont {Srolovitz}, \citenamefont
  {Ackland}, \citenamefont {Sun},\ and\ \citenamefont
  {Asta}}]{mendelevhan2003}%
  \BibitemOpen
  \bibfield  {author} {\bibinfo {author} {\bibfnamefont {M.~I.}\ \bibnamefont
  {Mendelev}}, \bibinfo {author} {\bibfnamefont {S.}~\bibnamefont {Han}},
  \bibinfo {author} {\bibfnamefont {D.~J.}\ \bibnamefont {Srolovitz}}, \bibinfo
  {author} {\bibfnamefont {G.~J.}\ \bibnamefont {Ackland}}, \bibinfo {author}
  {\bibfnamefont {D.~Y.}\ \bibnamefont {Sun}}, \ and\ \bibinfo {author}
  {\bibfnamefont {M.}~\bibnamefont {Asta}},\ }\href@noop {} {\bibfield
  {journal} {\bibinfo  {journal} {Philos. Mag.}\ }\textbf {\bibinfo {volume}
  {83}},\ \bibinfo {pages} {3977} (\bibinfo {year} {2003})}\BibitemShut
  {NoStop}%
\bibitem [{\citenamefont {Mills}\ and\ \citenamefont
  {J{\'{o}}nsson}(1994)}]{millsjonsson1994}%
  \BibitemOpen
  \bibfield  {author} {\bibinfo {author} {\bibfnamefont {G.}~\bibnamefont
  {Mills}}\ and\ \bibinfo {author} {\bibfnamefont {H.}~\bibnamefont
  {J{\'{o}}nsson}},\ }\href@noop {} {\bibfield  {journal} {\bibinfo  {journal}
  {Phy. Rev. Lett.}\ }\textbf {\bibinfo {volume} {72}},\ \bibinfo {pages}
  {1124} (\bibinfo {year} {1994})}\BibitemShut {NoStop}%
\bibitem [{\citenamefont {J{\'{o}}nsson}\ \emph {et~al.}(1998)\citenamefont
  {J{\'{o}}nsson}, \citenamefont {Mills},\ and\ \citenamefont
  {Jacobsen}}]{jonssonetal1998}%
  \BibitemOpen
  \bibfield  {author} {\bibinfo {author} {\bibfnamefont {H.}~\bibnamefont
  {J{\'{o}}nsson}}, \bibinfo {author} {\bibfnamefont {G.}~\bibnamefont
  {Mills}}, \ and\ \bibinfo {author} {\bibfnamefont {K.}~\bibnamefont
  {Jacobsen}},\ }\enquote {\bibinfo {title} {Classical and quantum dynamics in
  condensed phase simulations},}\ \ (\bibinfo  {publisher} {World Scientific},\
  \bibinfo {address} {Singapore},\ \bibinfo {year} {1998})\ Chap.~\bibinfo
  {chapter} {16}, pp.\ \bibinfo {pages} {385--404},\ \bibinfo {note} {edited by
  B. J. Berne, G. Cicotti, and D. F. Coker}\BibitemShut {NoStop}%
\bibitem [{\citenamefont {Henkelman}\ and\ \citenamefont
  {J{\'{o}}nsson}(2000)}]{henkelmanjonsson2000}%
  \BibitemOpen
  \bibfield  {author} {\bibinfo {author} {\bibfnamefont {G.}~\bibnamefont
  {Henkelman}}\ and\ \bibinfo {author} {\bibfnamefont {H.}~\bibnamefont
  {J{\'{o}}nsson}},\ }\href@noop {} {\bibfield  {journal} {\bibinfo  {journal}
  {J. Chem. Phys.}\ }\textbf {\bibinfo {volume} {113}},\ \bibinfo {pages}
  {9978} (\bibinfo {year} {2000})}\BibitemShut {NoStop}%
\bibitem [{\citenamefont {Einstein}(1905)}]{einstein1905}%
  \BibitemOpen
  \bibfield  {author} {\bibinfo {author} {\bibfnamefont {A.}~\bibnamefont
  {Einstein}},\ }\href@noop {} {\bibfield  {journal} {\bibinfo  {journal}
  {Annalen der Physik}\ }\textbf {\bibinfo {volume} {17}},\ \bibinfo {pages}
  {549} (\bibinfo {year} {1905})}\BibitemShut {NoStop}%
\bibitem [{\citenamefont {Langevin}(1908)}]{langevin1908}%
  \BibitemOpen
  \bibfield  {author} {\bibinfo {author} {\bibfnamefont {P.}~\bibnamefont
  {Langevin}},\ }\href@noop {} {\bibfield  {journal} {\bibinfo  {journal} {C.
  R. Acad. Sci. (Paris)}\ }\textbf {\bibinfo {volume} {146}},\ \bibinfo {pages}
  {530} (\bibinfo {year} {1908})}\BibitemShut {NoStop}%
\bibitem [{\citenamefont {Coffey}\ \emph {et~al.}(2004)\citenamefont {Coffey},
  \citenamefont {Kamkyov},\ and\ \citenamefont {Waldron}}]{coffeyetal2004}%
  \BibitemOpen
  \bibfield  {author} {\bibinfo {author} {\bibfnamefont {W.~T.}\ \bibnamefont
  {Coffey}}, \bibinfo {author} {\bibfnamefont {{\relax Yu}.~P.}\ \bibnamefont
  {Kamkyov}}, \ and\ \bibinfo {author} {\bibfnamefont {J.~T.}\ \bibnamefont
  {Waldron}},\ }\href@noop {} {\emph {\bibinfo {title} {The Langevin
  Equation}}},\ \bibinfo {edition} {2nd}\ ed.\ (\bibinfo  {publisher} {World
  Scientific, Singapore},\ \bibinfo {year} {2004})\BibitemShut {NoStop}%
\bibitem [{\citenamefont {Derlet}\ \emph {et~al.}(2007)\citenamefont {Derlet},
  \citenamefont {Nguyen-Manh},\ and\ \citenamefont
  {Dudarev}}]{derletmanhdudarev2007}%
  \BibitemOpen
  \bibfield  {author} {\bibinfo {author} {\bibfnamefont {P.~M.}\ \bibnamefont
  {Derlet}}, \bibinfo {author} {\bibfnamefont {D.}~\bibnamefont {Nguyen-Manh}},
  \ and\ \bibinfo {author} {\bibfnamefont {S.~L.}\ \bibnamefont {Dudarev}},\
  }\href@noop {} {\bibfield  {journal} {\bibinfo  {journal} {Phys. Rev. B}\
  }\textbf {\bibinfo {volume} {76}},\ \bibinfo {pages} {054107} (\bibinfo
  {year} {2007})}\BibitemShut {NoStop}%
\bibitem [{\citenamefont {Dudarev}\ \emph {et~al.}(2010)\citenamefont
  {Dudarev}, \citenamefont {Gilbert}, \citenamefont {Arakawa}, \citenamefont
  {Mori}, \citenamefont {Yao}, \citenamefont {Jenkins},\ and\ \citenamefont
  {Derlet}}]{dudarevgilbertetal2010}%
  \BibitemOpen
  \bibfield  {author} {\bibinfo {author} {\bibfnamefont {S.~L.}\ \bibnamefont
  {Dudarev}}, \bibinfo {author} {\bibfnamefont {M.~R.}\ \bibnamefont
  {Gilbert}}, \bibinfo {author} {\bibfnamefont {K.}~\bibnamefont {Arakawa}},
  \bibinfo {author} {\bibfnamefont {H.}~\bibnamefont {Mori}}, \bibinfo {author}
  {\bibfnamefont {Z.}~\bibnamefont {Yao}}, \bibinfo {author} {\bibfnamefont
  {M.~L.}\ \bibnamefont {Jenkins}}, \ and\ \bibinfo {author} {\bibfnamefont
  {P.~M.}\ \bibnamefont {Derlet}},\ }\href@noop {} {\bibfield  {journal}
  {\bibinfo  {journal} {Phys. Rev. B}\ }\textbf {\bibinfo {volume} {81}},\
  \bibinfo {pages} {224107} (\bibinfo {year} {2010})}\BibitemShut {NoStop}%
\bibitem [{\citenamefont {Frenkel}\ and\ \citenamefont
  {Smit}(2002)}]{frenkelsmit2002}%
  \BibitemOpen
  \bibfield  {author} {\bibinfo {author} {\bibfnamefont {D.}~\bibnamefont
  {Frenkel}}\ and\ \bibinfo {author} {\bibfnamefont {B.}~\bibnamefont {Smit}},\
  }\href@noop {} {\emph {\bibinfo {title} {Understanding Molecular Simulation:
  from Algorithms to Applications}}},\ \bibinfo {edition} {2nd}\ ed.\ (\bibinfo
   {publisher} {Academic Press, San Diego},\ \bibinfo {year}
  {2002})\BibitemShut {NoStop}%
\end{thebibliography}%

\end{document}